\documentclass{iau} 
\usepackage{graphicx}
\title[Synthesis of solid-state Complex Organic Molecules] 
{Synthesis of solid-state Complex Organic Molecules through accretion of simple species at low temperatures}

\author[D. Qasim]   
{D. Qasim$^1$,
G. Fedoseev$^2$, K.-J. Chuang$^{1,3}$\thanks{Present address: Laboratory Astrophysics Group of the Max Planck Institute for Astronomy at the Friedrich Schiller University Jena,
Institute of Solid State Physics, Helmholtzweg 3, D-07743 Jena, Germany}, V. Taquet$^4$, T. Lamberts$^5$, J. He$^1$, S. Ioppolo$^6$, E. F. van Dishoeck$^3$, 
 \and H. Linnartz{$^1$}}
 
\affiliation{$^1$Sackler Laboratory for Astrophysics, Leiden Observatory, Leiden University, PO Box 9513, NL--2300 RA Leiden, The Netherlands \\ email: {\tt dqasim@strw.leidenuniv.nl} \\[\affilskip]
{$^2$INAF--Osservatorio Astrofisico di Catania, via Santa Sofia 78, 95123 Catania, Italy}\\[\affilskip]{$^3$Leiden Observatory, Leiden University, PO Box 9513, NL--2300 RA Leiden, The Netherlands}\\[\affilskip]{$^4$INAF-Osservatorio Astrofisico di Arcetri, Largo E. Fermi 5, I-50125 Florence, Italy}\\[\affilskip]$^5$Leiden Institute of Chemistry, Leiden University, PO Box 9502, NL--2300 RA Leiden, The Netherlands\\[\affilskip]$^6$School of Electronic Engineering and Computer Science, Queen Mary University of London, Mile End Road, London E1 4NS, UK}

\pubyear{2019}
\volume{350}  
\jname{Laboratory Astrophysics: from Observations to Interpretation}
\editors{Farid Salama, Helen Jane Fraser, \& Harold Linnartz, eds.}
\begin{document}

\maketitle

\begin{abstract}

Complex organic molecules (COMs) have been detected in the gas-phase in cold and lightless molecular cores. Recent solid-state laboratory experiments have provided strong evidence that COMs can be formed on icy grains through `non-energetic' processes. In this contribution, we show that propanal and 1-propanol can be formed in this way at the low temperature of 10 K. Propanal has already been detected in space. 1-propanol is an astrobiologically relevant molecule, as it is a primary alcohol, and has not been astronomically detected. Propanal is the major product formed in the C$_2$H$_2$ + CO + H experiment, and 1-propanol is detected in the subsequent propanal + H experiment. The results are published in \cite[Qasim et al. (2019c)]{Qasim2019c}. ALMA observations towards IRAS 16293-2422B are discussed and provide a 1-propanol:propanal upper limit of $<  0.35-0.55$, which are complemented by computationally-derived activation barriers in addition to the performed laboratory experiments.      

\keywords{astrochemistry, astrobiology, methods: laboratory, telescopes, ISM: abundances, ISM: atoms, ISM: clouds, (ISM:) dust, extinction, ISM: molecules}
\end{abstract}

\firstsection 
\section{Introduction}
One of the earliest stages of the star formation cycle is the dense cloud stage. As the density increases to 10$^{4-5}$ cm$^{-3}$, it becomes increasingly difficult for external UV photons to penetrate the cloud, and also causes the temperature to decrease to about 10-20 K. From these conditions, it would appear that the level of chemical activity in these clouds is negligible. Yet, observations show that complex organic molecules are formed in these clouds and must have a solid-state origin (\cite[Soma et al. 2018]{Soma2018}).   

\begin{figure}[btp!]
\begin{center}
 \includegraphics[width=18cm]{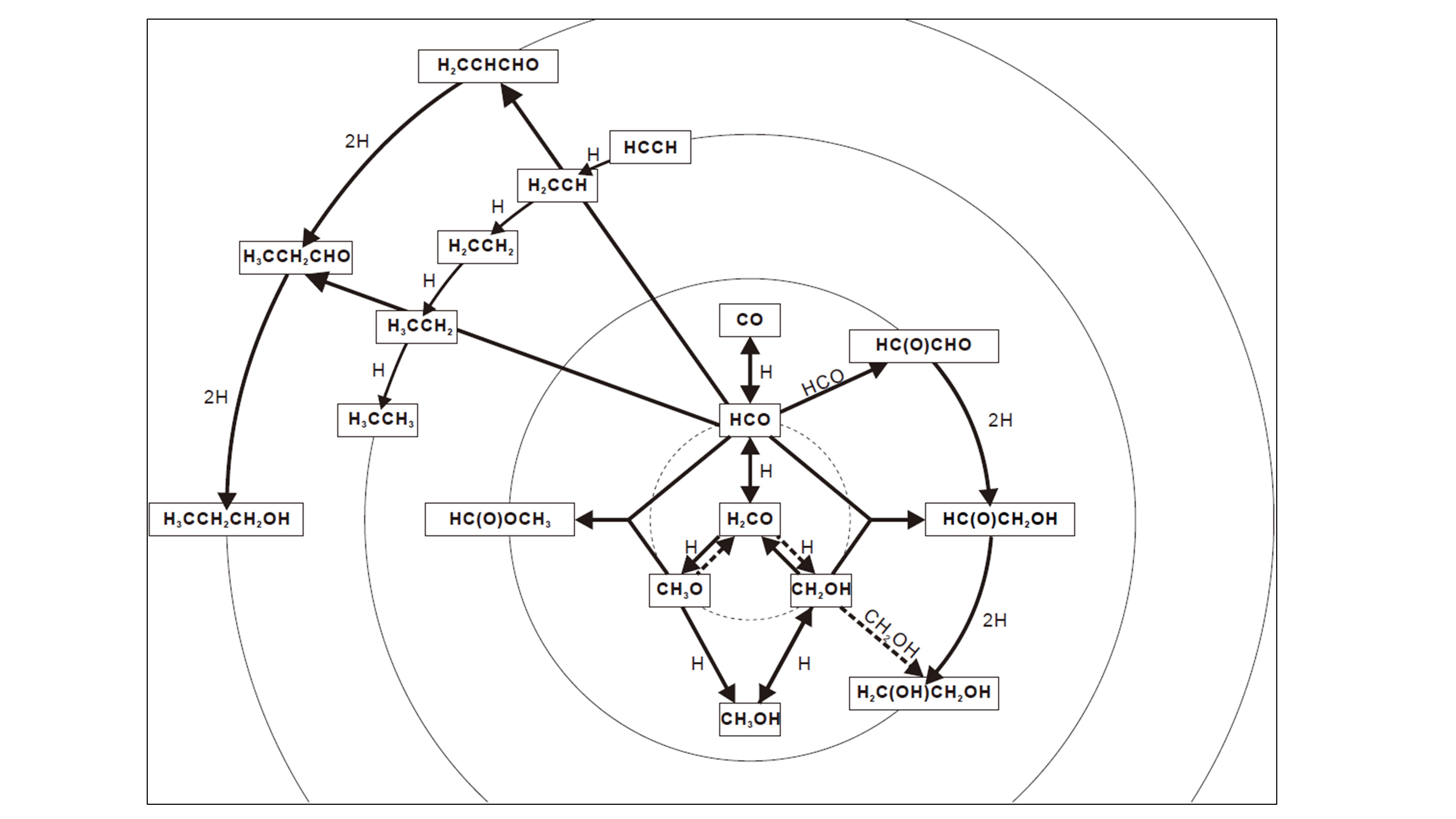} 
 \caption{Experimentally-derived COM formation network relevant to the CO freeze-out stage of dense clouds. Adapted from \cite[Chuang et al. (2016)]{Chuang2016}.}
   \label{fig1}
\end{center}
\end{figure}

The combination of observations (\cite[Boogert et al. 2015]{Boogert2015}), laboratory experiments (\cite[Watanabe et al. 2002]{Watanabe2002}; \cite[Fuchs et al. 2009]{Fuchs2009}), and computational simulations (\cite[Cuppen et al. 2007]{Cuppen2007}) shows that such dense clouds are further characterized by at least two distinct chemical phases: the H$_2$O-rich and CO-rich ice phases. Very recent laboratory experiments have shown that a number of alcohols have the potential to be formed during the H$_2$O-rich ice phase (\cite[Qasim et al. 2019a]{Qasim2019a}). COM formation relevant to the CO-rich ice phase has additionally been investigated, and includes the synthesis of glycolaldehyde, ethylene glycol, and methyl formate (\cite[Chuang et al. 2016]{Chuang2016}). This is displayed in the middle of Figure~\ref{fig1}. 

Here, we discuss the addition of hydrocarbon radicals to the CO + H reaction network to form the COMs, propanal and 1-propanol. The resulting extension of the reaction network is illustrated in Figure~\ref{fig1} and links to the well-studied CO hydrogenation chain. More details on this study, which also include ALMA observations and computationally-derived activation barriers, are found in \cite[Qasim et al. (2019c)]{Qasim2019c}.

\section{Experimental}

Experiments were performed with SURFRESIDE$^{2}$, an ultrahigh vacuum (UHV) apparatus, and details of the setup can be found in \cite[Ioppolo et al. (2013)]{Ioppolo2013}. On a 10 K surface, hydrogen atoms produced from a hydrogen atom beam source were used to hydrogenate C$_2$H$_2$ into C$_2$H$_x$ radicals and CO into CH$_x$O radicals, which can combine to form propanal and its hydrogenation product, 1-propanol. As 1-propanol was not clearly detected, also a propanal + H experiment was performed to unambiguously show 1-propanol formation. Detection techniques include temperature programmed desorption-quadrupole mass spectrometry (TPD-QMS) and reflection absorption infrared spectroscopy (RAIRS).      

\section{Results and discussion}

\begin{figure}[hbt!]
\begin{center}
 \includegraphics[width=10cm]{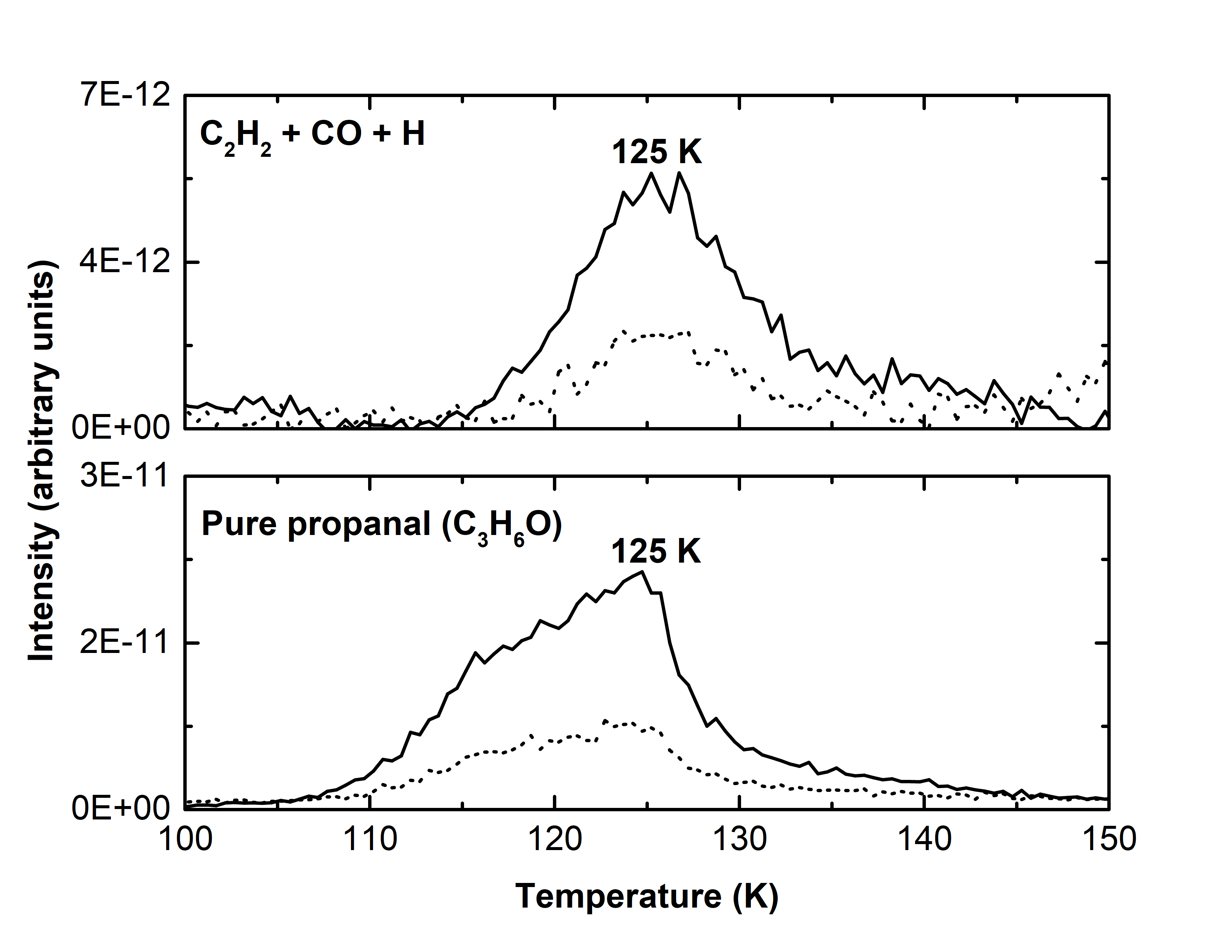} 
 \caption{Solid-line represents \emph{m/z} = 58, and dotted line represents \emph{m/z} = 57. (Top) TPD spectra of C$_2$H$_2$ + CO + H taken after deposition at 10 K. A 1:2:10 ratio of C$_2$H$_2$:CO:H was used. Flux values of $5 \times 10^{11}$ cm$^{-2}$ s$^{-1}$, $1 \times 10^{12}$ cm$^{-2}$ s$^{-1}$, and $5 \times 10^{12}$ cm$^{-2}$ s$^{-1}$ were used for C$_2$H$_2$, CO, and H, respectively (deposition time of 21600 seconds). (Bottom) TPD spectra of deposited propanal, with a propanal flux of $2 \times 10^{12}$ cm$^{-2}$ s$^{-1}$ (deposition time of 3600 seconds).}
   \label{fig2}
\end{center}
\end{figure}

\begin{figure}[hbt!]
\begin{center}
 \includegraphics[width=10cm]{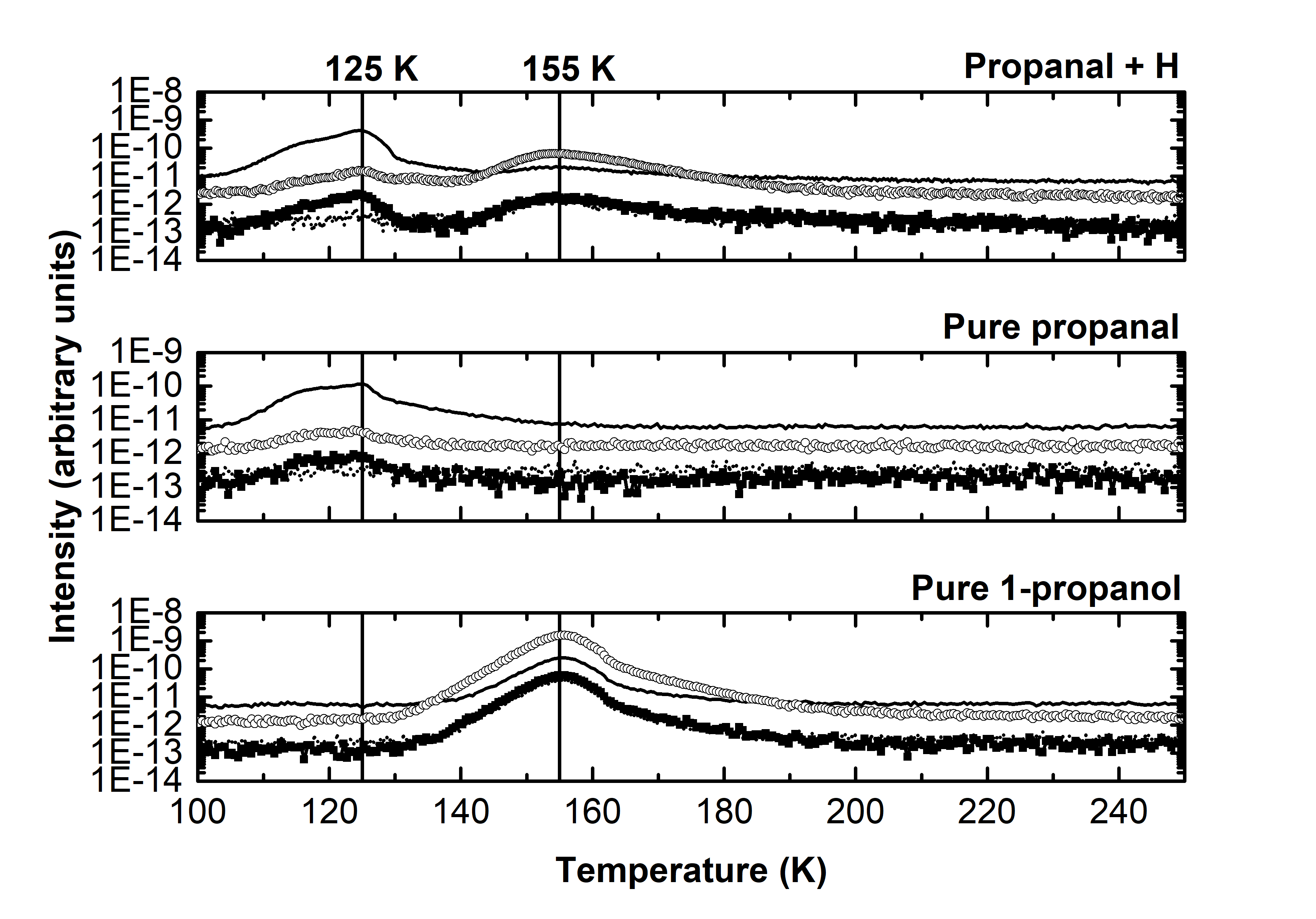} 
 \caption{Solid-line represents \emph{m/z} = 29, line with open-circle symbols represents \emph{m/z} = 31, line with filled-square symbols represents \emph{m/z} = 59, and dotted-line represents \emph{m/z} = 60. TPD spectra of propanal + H (top), propanal (middle), and 1-propanol (bottom) taken after deposition at 10 K. An H-flux of $5 \times 10^{12}$ cm$^{-2}$ s$^{-1}$ was used (deposition time of 7200 seconds). Propanal flux values of $2 \times 10^{11}$ cm$^{-2}$ s$^{-1}$ (deposition time of 7200 seconds; top) and $2 \times 10^{12}$ cm$^{-2}$ s$^{-1}$ (deposition time of 3600 seconds; middle) were used. 1-propanol flux of $1 \times 10^{12}$ cm$^{-2}$ s$^{-1}$ (deposition time of 3600 seconds; bottom) was used.}
   \label{fig3}
\end{center}
\end{figure}
Figure~\ref{fig2} shows the desorption of deposited propanal (bottom panel) and propanal formed in the C$_2$H$_2$ + CO + H experiment (top panel). Since the most abundant fragment of propanal is the parent ion, \emph{m/z} = 58 was probed, as well as the fragment, \emph{m/z} = 57, for extra confirmation. The peak desorption temperatures of 125 K are found in both experiments, which provides confirmation that propanal is formed in the C$_2$H$_2$ + CO + H experiment.

Figure~\ref{fig3} shows the desorption of deposited 1-propanol (bottom panel), deposited propanal (middle panel), and 1-propanol formed in the propanal + H experiment (top panel). The TPD data in the propanal + H experiment shows that \emph{m/z} values of 29, 31, 59, and 60 have peak desorptions at 155 K. This is also the peak desorption temperature of pure 1-propanol, as shown in the bottom panel, fully in-line with the idea that propanal is partially converted to 1-propanol upon hydrogenation. These results also agree with the computationally-derived activation barriers of $\sim$2500-2600 K from \cite[Zaverkin et al. (2018)]{Zaverkin2018}, in that the reaction can still proceed, although other possibly more efficient pathways to 1-propanol formation in the interstellar medium may exist, such as the pathways proposed in \cite[Qasim et al. (2019a)]{Qasim2019a}. The underlying chemical network is shown in Figure~\ref{fig1}. The important conclusion here is that the CO + H reaction chain, which is known to form many molecules identified in space already, additionally offers a starting point for other molecules to form, depending on the new radicals that are included.

\section{Astrochemical highlights and conclusions}
ALMA observations towards the low-mass protostar, IRAS 16293-2422B, yields a 1-propanol:propanal upper limit of $<  0.35-0.55$, where the 1-propanol column is from \cite[Qasim et al. (2019c)]{Qasim2019c} and the propanal column is from \cite[Lykke et al. (2017)]{Lykke2017}. By only taking into account the activation barrier(s) of propanal + H, it is expected that there is less 1-propanol in comparison to propanal if 1-propanol originates from propanal in those regions.     

In essence, solid-state propanal and 1-propanol have the potential to be formed in the CO freeze-out stage starting from CH$_x$O and C$_2$H$_x$ radicals. ALMA observations and computationally-derived activation barriers provide additional support that interstellar propanal can act as a precursor for 1-propanol. Although 1-propanol is currently undetected, its link to astrobiology, in that it is a primary alcohol, makes it a worthy target for future observational surveys. The chemical network presented here shows that such surveys should be performed in regions where propanal has been observed.    

\section{Acknowledgements}
This research is part of the Dutch Astrochemistry Network II (DANII).

\end{document}